\documentclass{article}
\usepackage{waspaa23,amsmath,graphicx,url,times}
\usepackage{makecell}
\usepackage{microtype}
\usepackage{color}
\usepackage{amsmath}
\usepackage{amssymb}
\usepackage{mathtools}
\usepackage{mathrsfs}
\usepackage{musicography}
\usepackage{tikz}
\usepackage{url}
\usetikzlibrary{positioning}
\usetikzlibrary{decorations.pathreplacing,calligraphy}

\DeclareMathOperator{\tr}{trace}

\newcommand\blfootnote[1]{%
  \begingroup
  \renewcommand\thefootnote{}\footnote{#1}%
  \addtocounter{footnote}{-1}%
  \endgroup
}

\title{FITTING AUDITORY FILTERBANKS WITH MULTIRESOLUTION NEURAL NETWORKS}

\name{Vincent Lostanlen$^{1}$,
      Daniel Haider$^{2,3}$,
      Han Han$^{1}$,
      Mathieu Lagrange$^{1}$,
      Peter Balazs$^{2}$,
      and Martin Ehler$^{3}$}
\address{$^1$ Nantes Université, École Centrale Nantes, CNRS, LS2N, UMR 6004, F-44000 Nantes, France.\\
         $^2$ Acoustics Research Institute, Austrian Academy of Sciences, A-1040 Vienna, Austria.\\
         $^3$ University of
Vienna, Department of Mathematics, A-1090 Vienna, Austria.
}

\begin{document}

\ninept
\maketitle

\begin{sloppy}

\begin{abstract}
Waveform-based deep learning faces a dilemma between nonparametric and parametric approaches.
On one hand, convolutional neural networks (convnets) may approximate any linear time-invariant system; yet, in practice, their frequency responses become more irregular as their receptive fields grow.
On the other hand, a parametric model such as LEAF is guaranteed to yield Gabor filters, hence an optimal time--frequency localization; yet, this strong inductive bias comes at the detriment of representational capacity.
In this paper, we aim to overcome this dilemma by introducing a neural audio model, named multiresolution neural network (MuReNN).
The key idea behind MuReNN is to train separate convolutional operators over the octave subbands of a discrete wavelet transform (DWT).
Since the scale of DWT atoms grows exponentially between octaves, the receptive fields of the subsequent learnable convolutions in MuReNN are dilated accordingly.
For a given real-world dataset, we fit the magnitude response of MuReNN to that of a well-established auditory filterbank: Gammatone for speech, CQT for music, and third-octave for urban sounds, respectively.
This is a form of knowledge distillation (KD), in which the filterbank ``teacher'' is engineered by domain knowledge while the neural network ``student'' is optimized from data.
We compare MuReNN to the state of the art in terms of goodness of fit after KD on a hold-out set and in terms of Heisenberg time--frequency localization.
Compared to convnets and Gabor convolutions, we find that MuReNN reaches state-of-the-art performance on all three optimization problems.
\end{abstract}


\section{Introduction}
\label{sec:intro}
Auditory filterbanks are time-invariant systems whose design takes inspiration from domain-specific knowledge in hearing science \cite{lyon2017human}.
For example, the critical bands of the human cochlea inspires frequency scales such as mel, bark, and ERB \cite{knight2021cambridge}.
The phenomenon of temporal masking calls for asymmetric impulse responses, motivating the design of Gammatone filters \cite{Glasberg1990gammatone}.
Lastly, the constant-$Q$ transform (CQT), in which the number of filters per octave is fixed, reflects the principle of octave equivalence in music \cite{brown1991calculation}. 

In recent years, the growing interest for deep learning in signal processing has proposed to learn filterbanks from data rather than design them a priori \cite{dorfler2020basic}.
Such a replacement of feature engineering to feature learning is motivated by the diverse application scope of audio content analysis: i.e., conservation biology \cite{stowell2022computational}, urban science \cite{bello2019sonyc}, industry \cite{zhao2019deep}, and healthcare \cite{bizopoulos2018deep}.
Since these applications differ greatly in terms of acoustical content, the domain knowledge which prevails in speech and music processing is likely to yield suboptimal performance.
Instead, gradient-based optimization has the potential to reflect the spectrotemporal characteristics of the data at hand.

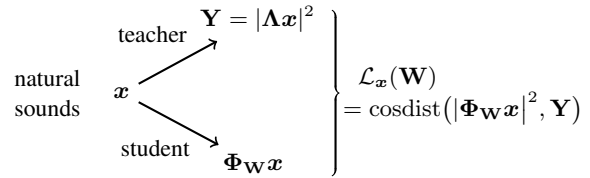
\begin{figure}
\centering
\begin{tikzpicture}[thick]
\node(x){$\boldsymbol{x}$};
\node(natural)[left=of x, yshift=2mm, xshift=8mm]{natural};
\node(sounds)[left=of x, yshift=-2mm, xshift=8mm]{sounds};
\node(Sx)[above right=of x,yshift=-5mm,xshift=-3mm]{$\mathbf{Y}=\vert\boldsymbol{\Lambda}\boldsymbol{x}\vert^2$};
\node(yf)[below right=of x,yshift=5mm]{$\boldsymbol{\Phi}_\mathbf{W}\boldsymbol{x}$};
\draw[->] (x) -- (Sx) node[xshift=-14mm,yshift=-2mm]{teacher};
\draw[->] (x) -- (yf) node[xshift=-13mm,yshift=2mm]{student};
\draw[decorate, decoration = {calligraphic brace, mirror}] (2.8,-1.1) --  (2.8,1.1) node(L)[right=of x, xshift=18mm, yshift=2mm]{$\displaystyle{\mathcal{L}_{\boldsymbol{x}}(\mathbf{W})}$};
\node(Lrhs)[right=of x, xshift=16mm, yshift=-2mm]{$\displaystyle{=\mathrm{cosdist}\big(\vert\mathbf{\Phi_{\mathbf{W}}}\boldsymbol{x}\big\vert^2, \mathbf{Y}\big)}$};
\end{tikzpicture}
\caption{Graphical outline of the proposed method.
We train a neural network ``student'' $\boldsymbol{\Phi}_{\mathbf{W}}$ to regress the squared magnitudes $\mathbf{Y}$ of an auditory filterbank ``teacher'' $\boldsymbol{\Lambda}$ in terms of spectrogram-based cosine distance $\mathcal{L}_{\mathbf{x}}$, on average over a dataset of natural sounds $\boldsymbol{x}$.
\label{fig:overview}}
\end{figure}

Enabling this potential is particularly important in applications where psychoacoustic knowledge is lacking; e.g., animals outside of the mammalian taxon \cite{bravo2021bioacoustic,faiss2022adaptive}.
Beyond its perspectives in applied science, the study of learnable filterbanks has value for fundamental research on machine listening with AI.
This is because it represents the last stage of progress towards general-purpose ``end-to-end'' learning, from the raw audio waveform to the latent space of interest.

Yet, success stories in waveform-based deep learning for audio classification have been, up to date, surprisingly few---and even fewer beyond the realm of speech and music \cite{lluis2018end}.
The core hypothesis of our paper is that this shortcoming is due to an inadequate choice of neural network architecture.
Specifically, we identify a dilemma between nonparametric and parametric approaches, where the former are represented by convolutional neural networks (convnets) and the latter by architectures used in SincNet \cite{ravanelli2018speaker} or LEAF \cite{zeghidour2021leaf}.
In theory, convnets may approximate any finite impulse response (FIR), given a receptive field that is wide enough; but in practice, gradient-based optimization on nonconvex objectives yields suboptimal solutions \cite{lluis2018end}.
On the other hand, the parametric approaches enforce good time--frequency localization, yet at the cost of imposing a rigid shape for the learned filters: cardinal sine (inverse-square envelope) for SincNet and Gabor (Gaussian envelope) for LEAF.

Our goal is to overcome this dilemma by developing a neural audio model which is capable of learning temporal envelopes from data while guaranteeing near-optimal time--frequency localization\blfootnote{Companion website: \url{https://github.com/lostanlen/lostanlen2023waspaa}}.
In doing so, we aim to bypass the explicit incorporation of psychoacoustic knowledge as much as possible.
This is unlike state-of-the-art convnets for filterbank learning such as SincNet or LEAF, whose parametric kernels are initialized according to a mel-frequency scale.
Arguably, such careful initialization procedures defeat the purpose of deep learning; i.e., to spare the human effort of feature engineering.
Furthermore, it contrasts with other domains of deep learning (e.g., image processing) in which all convnet layers are simply initialized with i.i.d. Gaussian weights \cite{sutskever2013importance}.

Prior work on this problem has focused on advancing the state of the art on a given task, sometimes to no avail  \cite{schluter2022efficientleaf}.
In this article, we take a step back and formulate a different question: before we try to outperform an auditory filterbank, can we replicate its responses with a neural audio model?
To answer this question, we compare different ``student'' models in terms of their ability to learn from a black-box function or ``teacher'' by knowledge distillation (KD).

Given an auditory filterbank $\boldsymbol{\Lambda}$ and a discrete-time signal $\boldsymbol{x}$ of length $T$, let us denote the squared magnitude of the filter response at frequency bin $f$ by $\boldsymbol{Y}[f,t] = \vert \boldsymbol{\Lambda}\boldsymbol{x} \vert^2 [f, 2^J t]$, where $2^J$ is the chosen hop size or ``stride''.
Then, given a model $\Phi_{\mathbf{W}}$ with weights $\mathbf{W}$, we evaluate the dissimilarity between teacher $\boldsymbol{\Lambda}$ and student $\mathbf{\Phi_{\mathbf{W}}}$ as their (squared) spectrogram-based cosine similarity $\mathcal{L}_{\boldsymbol{x}}(\mathbf{W})$. 
The distance of student and teacher in this similarity measure can be computed via the $L^2$ distance after normalizing across frequency bins $f$, independently for each time $t$. Let $\vert\boldsymbol{\widetilde{\Phi}}_{\mathbf{W}}\boldsymbol{x}\big\vert^2$ and $\mathbf{\widetilde{Y}}$ denote these normalized versions of student and teacher, then
\begin{align}
\mathcal{L}_{\boldsymbol{x}}(\mathbf{W}) &=
\mathrm{cosdist}\big(\vert\boldsymbol{\Phi}_{\mathbf{W}}\vert^2, \mathbf{Y}\big) \nonumber \\
&= \dfrac{1}{2} \sum_{t=1}^{T/2^J} \sum_{f=1}^{F}
\big\vert \vert\boldsymbol{\widetilde{\Phi}}_{\mathbf{W}}\boldsymbol{x}\big\vert^2[f,t] - \mathbf{\widetilde{Y}}[f,t] \big\vert^2,
\label{eq:loss}
\end{align}
where $F$ is the number of filters.
We seek to minimize the quantity above by gradient-based optimization on $\mathbf{W}$, on a real-world dataset of audio signals $\{\boldsymbol{x_1} \ldots \boldsymbol{x_N}\}$, and with no prior knowledge on $\boldsymbol{\Lambda}$.

\section{Neural audio models}

\subsection{Learnable time-domain filterbanks (Conv1D)}
As a baseline, we train a 1-D convnet $\mathbf{\Phi_{\mathbf{W}}}$ with $F$ kernels of the same length $2L$.
With a constant stride of $2^J$, $\mathbf{\Phi_{\mathbf{W}}}\boldsymbol{x}$ writes as
\begin{equation}
\mathbf{\Phi_{\mathbf{W}}}\boldsymbol{x}[f,t] =
(\boldsymbol{x}\ast\boldsymbol{\phi_f})[2^J t] =
\sum_{\tau=-L}^{L-1} \boldsymbol{x}\big[2^J t-\tau\big]
\boldsymbol{\phi_f}[\tau],
\label{eq:fir}
\end{equation}
where $\boldsymbol{x}$ is padded by $L$ samples at both ends.
Under this setting, the trainable weights ${\mathbf{W}}$ are the finite impulse responses of $\boldsymbol{\phi_f}$ for all $f$, thus amounting to $2LF$ parameters.We initialize $\mathbf{W}$ as Gaussian i.i.d. entries with null mean and variance $1/\sqrt{F}$.

\subsection{Gabor 1-D convolutions (Gabor1D)}
As a representative of the state of the art (i.e., LEAF \cite{zeghidour2021leaf}), we train a Gabor filtering layer or Gabor1D for short.
For this purpose, we parametrize each FIR filter $\boldsymbol{\phi_f}$ as Gabor filter; i.e., an exponential sine wave of amplitude $a_f$ and frequency $\eta_f$ which is modulated by a Gaussian envelope of width $\sigma_f$.
Hence a new definition:
\begin{equation}
\boldsymbol{\phi_f}[\tau] = 
\dfrac{a_f}{\sqrt{2\pi}\sigma_f}
\exp\left(-
\dfrac{\tau^2}{2\sigma_f^2}
\right)
\exp(2\pi\mathrm{i}\eta_f \tau).
\end{equation}

Under this setting, the trainable weights $\mathbf{W}$ amount to only $3F$ parameters:
${\mathbf{W}} = \{a_1, \sigma_1, \eta_1, \ldots, a_F, \sigma_F, \eta_F \}$.
Following LEAF, we initialize center frequencies $\eta_f$ and bandwidths $\sigma_f$ so as to form a mel-frequency filterbank \cite{zeghidour2018learning} and set amplitudes $a_f$ to one.
We use the implementation of Gabor1D from SpeechBrain v0.5.14 \cite{ravanelli2021speechbrain}.

\subsection{Multiresolution neural network (MuReNN)}
As our original contribution, we train a multiresolution neural network, or MuReNN for short.
MuReNN comprises two stages, multiresolution approximation (MRA) and convnet; of which only the latter is learned from data.
We implement the MRA with a dual-tree complex wavelet transform (DTCWT) \cite{selesnick2005dual}.
The DTCWT relies on a multirate filterbank in which each wavelet $\boldsymbol{\psi_j}$ has a null average and a bandwidth of one octave.
Denoting by $\xi$ the sampling rate of $\boldsymbol{x}$, the wavelet $\boldsymbol{\psi_j}$ has a bandwidth with cutoff frequencies $2^{-(j+1)}\pi$ and $2^{-j}\pi$.
Hence, we may subsample the result of the convolution $(\boldsymbol{x} \ast \boldsymbol{\psi_j})$ by a factor of $2^j$, yielding:
\begin{equation}
\forall j\in\{0,\ldots,J-1\},\;\boldsymbol{x_{j}}[t] =
(\boldsymbol{x} \ast \boldsymbol{\psi_j})[2^j t],
\label{eq:mra}
\end{equation}
where $J$ is the number of multiresolution levels.
We take $J=9$ in this paper, which roughly coincides with the number of octaves in the hearing range of humans.
The second stage in MuReNN consists in defining convnet filters $\boldsymbol{\phi_f}$.
Unlike in the Conv1D setting, those filters do not operate over the full-resolution input $\boldsymbol{x}$ but over one of its MRA levels $\boldsymbol{x_{j}}$.
More precisely, let us denote by $j[f]$ the decomposition level assigned to filter $f$, and by $2L_j$ the kernel size for that decomposition level.
We convolve $\boldsymbol{x}_{j[f]}$ with $\boldsymbol{\phi_f}$ and apply a subsampling factor of $2^{J-j[f]}$, hence:
\begin{align}
\mathbf{\Phi_W}\boldsymbol{x}[f,t] &=
(\boldsymbol{x_{j[f]}}\ast\boldsymbol{\phi_f})[2^{J-j[f]} t] \nonumber \\
&= \sum_{\tau=-L_j}^{L_j-1} \boldsymbol{x_{j[f]}}\big[2^{J-j[f]} t-\tau\big]
\boldsymbol{\phi_f}[\tau]
\label{eq:murenn}
\end{align}
The two stages of subsampling in Equations \ref{eq:mra} and \ref{eq:murenn} result in a uniform downsampling factor of $2^J$ for $\mathbf{\Phi_W}\boldsymbol{x}$.
Each learned FIR filter $\boldsymbol{\phi}_f$ has an effective receptive field size of $2^{j[f]+1}L_{j[f]}$, thanks to the subsampling operation in Equation \ref{eq:mra}.
This resembles a dilated convolution \cite{oord2018parallel} with a dilation factor of $2^{j[f]}$, except that the DTCWT guarantees the absence of aliasing artifacts.

Besides this gain in frugality, as measured by parameter count per unit of time, the resort to an MRA offers the opportunity to introduce desirable mathematical properties in the non-learned part of the transform (namely, $\boldsymbol{\psi_f}$) and have the MuReNN operator $\boldsymbol{\Phi}_{\mathbf{W}}$ inherit them, without need for a non-random initialization nor regularization during training.
In particular, $\boldsymbol{\Phi}_{\mathbf{W}}$ has at least as many vanishing moments as $\boldsymbol{\psi}_f$.
Furthermore, the DTCWT yields quasi-analytic coefficients: for each $j$, $\boldsymbol{x_{j}} = \boldsymbol{x}^{\mathbb{R}}_{\boldsymbol{j}} + \mathrm{i} \boldsymbol{x}^{\mathbb{I}}_{\boldsymbol{j}}$ with $\boldsymbol{x}^{\mathbb{I}}_{\boldsymbol{j}} \approx \mathcal{H}\left(\boldsymbol{x}^{\mathbb{R}}_{\boldsymbol{j}}\right)$, where the exponent $\mathbb{R}$ (resp. $\mathbb{I}$) denotes the real part (resp. imaginary part) and $\mathcal{H}$ denotes the Hilbert transform.
Since $\boldsymbol{\phi_f}$ is real-valued, the same property holds for MuReNN: $\mathbf{\Phi}^{\mathbb{I}}\boldsymbol{x} = \mathcal{H}(\mathbf{\Phi}^{\mathbb{R}}\boldsymbol{x})$.

We implement MuReNN on GPU via a custom implementation of DTCWT in PyTorch\footnote{\url{https://github.com/kymatio/murenn}}.
Following \cite{selesnick2005dual}, we use a biorthogonal wavelet for $j=0$ and quarter-shift wavelets for $j\geq1$.
We set $L_j = 8M_j$ where $M_j$ is the number of filters $f$ at resolution $j$.
We refer to \cite{cotter2020uses} for an introduction to deep learning in the wavelet domain, with applications to image classification.



\begin{table*}[!ht]
    \centering
    \begin{tabular}{lll||rrr}
    \makecell{Domain} & \makecell{Dataset} & \makecell{Teacher}
    &\makecell{Conv1D} & \makecell{Gabor1D} & \makecell{MuReNN} \\
    \hline
    Speech & NTVOW & Gammatone & $2.12 \pm 0.05$ & $10.14 \pm 0.09$ & $\mathbf{2.00 \pm 0.02}$ \\
    Music & TinySOL & VQT & $8.76 \pm 0.2$ & $16.87 \pm 0.06$ & $\mathbf{5.28 \pm 0.03}$ \\
    Urban & SONYC-UST & ANSI S1.11 & $3.26 \pm 0.1$ & $13.51 \pm 0.2$ & $\mathbf{2.57 \pm 0.2}$\\
    Synth & Sine waves & CQT & $11.54 \pm 0.5$ & $22.26 \pm 0.9$ & $\mathbf{9.75 \pm 0.4}$
    \end{tabular}
    \caption{
    Mean and standard deviation of test loss after knowledge distillation over five independent trials.
    Each column corresponds to a different neural audio model $\boldsymbol{\Phi}_{\mathbf{W}}$ while each row corresponds to a different auditory filterbank and audio domain.
    See Section \ref{sub:benchmark} for details.
    }
    \label{tab:results}
\end{table*}

\begin{figure*}
  \centering
  \centerline{
  \includegraphics[width=\linewidth]{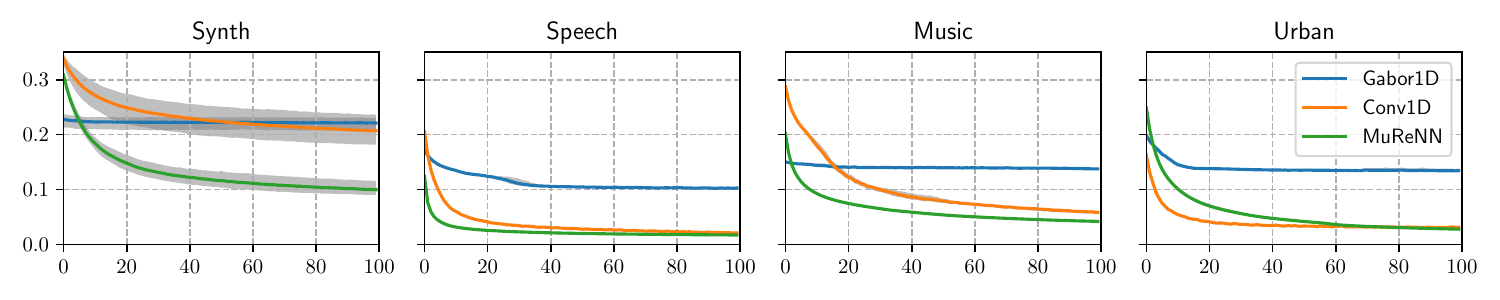}
  }
  \caption{
  Left to right: evolution of validation losses on different domains with Conv1D (green), Gabor1D (blue), and MuReNN (orange), as a function of training epochs.
  The shaded area denotes the standard deviation across five independent trials.
  See Section \ref{sub:benchmark} for details.
  }
  \label{fig:convergence}
\end{figure*}

\section{Knowledge distillation}
\subsection{Target auditory filterbanks}\label{sec:teacher}
For each of the three different domains, speech, music and urban environmental sounds, we use an auditory filterbank $\boldsymbol{\Lambda}$ that is tailored to its respective spectrotemporal characteristics.

\begin{description}
\item[Synth]
A constant-$Q$ filterbank with $Q=8$ filters per octave, covering eight octaves with Hann-modulated sine waves.
\item[Speech]
A filterbank with 4-th order Gammatone filters tuned to the ERB-scale, a frequency scale which is adapted to the equivalent rectangular bandwidths of the human cochlea \cite{moore1983erb}.
In psychoacoustics, Gammatone filters provide a good approximation to measured responses of the filters of the human basilar membrane \cite{Glasberg1990gammatone}.
Unlike Gabor filters, Gammatone filters are asymmetric, both in the time domain and frequency domain.
We refer to \cite{necciari2018audlets} for implementation details.

\item[Music]
A variable-$Q$ transform (VQT) with $M_j = 12$ frequency bins per octave at every level.
The VQT is a variant of the constant-$Q$ transform (CQT) in which $Q$ is decreased gradually towards lower frequencies \cite{schörkhuber2014vqt}, hence an improved temporal resolution at the expense of frequency resolution.

\item[Urban]
A third-octave filterbank inspired by the ANSI S1.11-2004 standard for environmental noise monitoring \cite{antoni2010orthogonal}.
In this filterbank, center frequencies are not exactly in a geometric progression.
Rather, they are aligned with integer Hertz values: 40, 50, 60; 80, 100, 120; 160, 200, 240; and so forth.
\end{description}
We construct the Synth teacher via nnAudio \cite{cheuk2020nnaudio}, a PyTorch port of librosa \cite{mcfee23librosa}; and Speech, Music, and Urban using the Large Time--Frequency Analysis Toolbox (LTFAT) for MATLAB \cite{pruvsa2013ltfat}.

\subsection{Gradient-based optimization}
For all four ``student'' models, we initialize the vector $\mathbf{W}$ at random and update it iteratively by empirical risk minimization over the training set. 
We rely on the Adam algorithm for stochastic optimization with default momentum parameters.
Given the definition of spectrogram-based cosine distance in Equation \ref{eq:loss}, we perform reverse-mode automatic differentiation in PyTorch to obtain
\begin{align}
\boldsymbol{\nabla}
\mathcal{L}_{\boldsymbol{x}}(\mathbf{W})[i] =
\sum_{f=1}^{F} \sum_{t=1}^{T/2^J}
&\dfrac{\partial \vert\mathbf{\widetilde{\Phi}_{\mathbf{W}}}\boldsymbol{x}\big\vert^2[f,t]}{\partial \mathbf{W}[i]}(\mathbf{W}) \nonumber \\
&\times \big( \vert\mathbf{\widetilde{\Phi}_{\mathbf{W}}}\boldsymbol{x}\big\vert^2[f,t] - \mathbf{\widetilde{Y}}[f,t] \big)
\label{eq:grad}
\end{align}
for each entry $\mathbf{W}[i]$.
Note that the gradient above does not involve the phases of the teacher filterbank $\boldsymbol{\Lambda}$, only its normalized magnitude response $\boldsymbol{\mathbf{Y}}$ given the input $\boldsymbol{x}$.
Consequently, even though our models $\mathbf{\Phi}_\mathbf{W}$ contain a single linear layer, the associated knowledge distillation procedure is nonconvex, and thus resembles the training of a deep neural network.

\section{Results and discussion}
\subsection{Datasets}
\begin{description}
    \item[Synth] As a proof of concept, we construct sine waves in a geometric progression over the frequency range of the target filterbank.
    \item[Speech] The North Texas vowel database (NTVOW) \cite{assmann2000jasa} contains utterances of 12 English vowels from 50 American speakers, including children aged three to seven as well as male and female adults. In total, it consists of 3190 recordings, each lasting between one and three seconds.
    \item[Music] The TinySOL dataset \cite{cella2020icmc} contains isolated musical notes played by eight instruments: accordion, alto saxophone, bassoon, flute, harp, trumpet in C, and cello. For each of these instruments, we take all available pitches in the tessitura (min = $B_0$, median = $E_4$, max = $C\sharp_8$ ) in three levels of intensity dynamics: \textit{pp}, \textit{mf}, and \textit{ff}. This results in a total of 1212 audio recordings.
    \item[Urban] The SONYC Urban Sound Tagging dataset (SONYC-UST) \cite{cartwright2019sonyc} contains 2803 acoustic scenes from a network of autonomous sensors in New York City. Each of these ten-second scenes contains one or several sources of urban noise pollution, such as: engines, machinery and non-machinery impacts, powered saws, alert signals, and dog barks.
\end{description}

\begin{figure}[t]
  \centering
  \centerline{
  \includegraphics[width=\columnwidth]{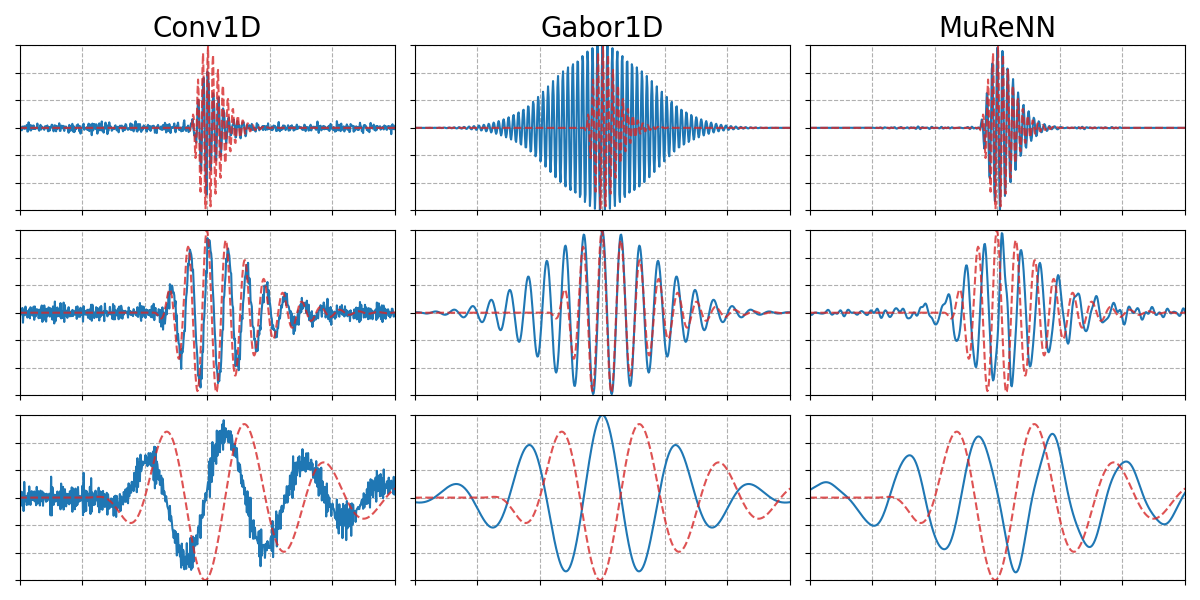}}
  \caption{
  Compared impulse responses of Conv1D (left), Gabor1D (center), and MuReNN (right) with different center frequencies after convergence, with a Gammatone filterbank as target.
  Solid blue (resp. dashed red) lines denote the real part of the impulse responses of the learned filters (resp. target).
  See Section \ref{sub:qualitative} for details.
  }
  \label{fig:impulse-responses}
\end{figure}

\subsection{Benchmarks}
\label{sub:benchmark}
For each audio domain, we randomly split its corresponding dataset into training, testing and validation subsets with a 8:1:1 ratio.
During training, we select $2^{12}$ time samples from the middle part of each signal, i.e., the FIR length of the filters in the teacher filterbank. 
We train each model with 100 epochs with an epoch size of 8000.

Table \ref{tab:results} summarizes our findings.
On all three benchmarks, we observe that MuReNN reaches state-of-the-art performance, as measured in terms of cosine distance with respect to the teacher filterbank after 100 epochs.
The improvement with respect to Conv1D is most noticeable in the Synth benchmark and least noticeable in the Speech benchmark.
Furthermore, Figure \ref{fig:convergence} indicates that Gabor1D barely trains at all: this observation is consistent with the sensitivity of LEAF with respect to initialization, as reported in \cite{anderson2023learnable}.
We also notice that MuReNN trains faster than Conv1D on all benchmarks except for Urban, a phenomenon deserving further inquiry.

\subsection{Error analysis}
\label{sub:qualitative}
The mel-scale initialization of Gabor1D filters and the inductive bias of MuReNN enabled by octave localization gives a starting advantage when learning filterbanks on log-based frequency scales, as used for the Gammatone and VQT filterbank.
Expectedly, this advantage is absent with a teacher filterbank that does not follow a geometric progression of center frequencies, as it is the case in the ANSI scale. Figure \ref{fig:convergence} reflects these observations.

To examine the individual filters of each model, we take the speech domain as an example and obtain their learned impulse responses.
Figure \ref{fig:impulse-responses} visualizes chosen examples at different frequencies learned by each model together with the corresponding teacher Gammatone filters.
In general, all models are able to fit the filter responses well.
However, it is noticeable that the prescribed envelope for Gabor1D impedes it from learning the asymmetric target Gammatone filters. This becomes prominent especially at high frequencies.
From the strong envelope mismatches at coinciding frequency
we may deduce that center frequencies and bandwidths did not play well together during training.
On the contrary, MuReNN and Conv1D are flexible enough to learn asymmetric temporal envelopes without compromising its regularity in time.
Although the learned filters of Conv1D are capable of fitting the frequencies well, they suffer from noisy artifacts, especially outside their essential supports.
Indeed, through limiting the scale and support of the learned filters, MuReNN restrains the potential introduction of high-frequency noises of a learned filter of longer length.
The phase misalignment at low frequencies is a natural consequence of the fact that the gradients are computed from the magnitudes of the filterbank responses.

Finally, we measure the time--frequency localization of all filters by computing the associated Heisenberg time--frequency ratios \cite{mallat1999wavelet}. From theory we know that Gaussian windows are optimal in this sense \cite{gröchenig01tfana}.
Therefore, it is not surprising that Gabor1D yields the best localized filters, even outperforming the teacher, see Figure \ref{fig:heisenberg}.
Expectedly, the localization of the filters from Conv1D is poor and appears independent of the teacher. MuReNN roughly resembles the localization of the teachers but has some poorly localized outliers in higher frequencies, deserving further inquiry.

\begin{figure}[t]
  \centering
  \centerline{
  \includegraphics[width=0.8\columnwidth]{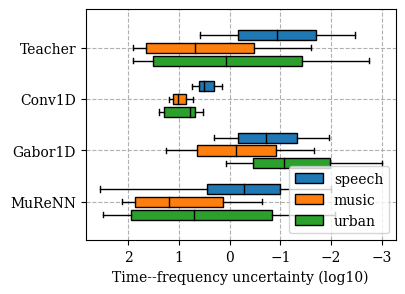}}
  \caption{Distribution of Heisenberg time--frequency ratios for each teacher--student pair (lower is better).
  See Section \ref{sub:qualitative} for details.}
  \label{fig:heisenberg}
\end{figure}

\section{Conclusion}
Multiresolution neural networks (MuReNN) have the potential to advance waveform-based deep learning.
They offer a flexible and data-driven procedure for learning filters which are ``wavelet-like'': i.e., narrowband with compact support, vanishing moments, and quasi-Hilbert analyticity.
Those experiments based on knowledge distillation from three domains (speech, music, and urban sounds) illustrate the suitability of MuReNN for real-world applications.
The main limitation of MuReNN lies in the need to specify a number of filters per octave $M_j$, together with a kernel size $L_j$.
Still, a promising finding of our paper is that prior knowledge on $M_j$ and $L_j$ suffices to finely approximate non-Gabor auditory filterbanks, such as Gammatones on an ERB scale, from a random i.i.d. Gaussian initialization.
Future work will evaluate MuReNN in conjunction with a deep neural network for sample-efficient audio classification.





\section{ACKNOWLEDGMENT}
\label{sec:ack}

V.L. thanks Fergal Cotter and Nick Kingsbury for maintaining the dtcwt and pytorch\_wavelets libraries; LS2N and ÖAW staff for arranging research visits; and Neil Zeghidour for helpful discussions. D.H. thanks Clara Hollomey for helping with the implementation of the filterbanks.
V.L. and M.L. are supported by ANR MuReNN; D.H., by a DOC Fellowship of the Austrian Academy of Sciences (A 26355); P.B., by FWF projects LoFT (P 34624) and NoMASP (P 34922); and M.E., by WWTF project CHARMED (VRG12-009).

\newpage
\newpage

\bibliographystyle{IEEEtran}
\bibliography{refs23}

\end{sloppy}
\end{document}